\documentclass[preprint,5p,12pt]{elsarticle}

\usepackage{graphicx}
\usepackage{dcolumn}
\usepackage{bm}
\usepackage{amssymb}
\usepackage{amsmath}
\usepackage{color}
\usepackage{physics}
\biboptions{sort&compress}
\usepackage{soul}
\usepackage{hyperref}
\usepackage{microtype}
\usepackage[normalem]{ulem}

\hypersetup{
    colorlinks=true, linkcolor=blue, citecolor=blue,
    filecolor=blue, urlcolor=blue, breaklinks=true
}

\newcommand{\orcid}[1]{\href{https://orcid.org/#1}
  {\includegraphics[width=7pt]{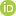}}}

\journal{arXiv}

\begin{document}

\begin{frontmatter}

\title{Stochastic reversal of deterministic selection in epidemic strain competition}

\author{Enrique C. Gabrick$^{1*}$\orcid{0000-0002-8407-7675}}
\author{Ana Luiza de Moraes$^{1}$\orcid{0000-0003-0565-4620}}
\author{Ervin K. Lenzi$^{2}$\orcid{0000-0003-3853-1790}}
\author{Iberê L. Caldas$^{1}$\orcid{0000-0003-3853-1790}}

\address{
$^1$Institute of Physics, University of São Paulo, 
05508-090 São Paulo, São Paulo, Brazil.\\
$^2$Department of Physics, State University of Maring\'a, 87020-900, Maring\'a, Paraná, Brazil.}

\cortext[cor]{ecgabrick@gmail.com}

\begin{abstract}
Different strains competing for a common pool of susceptible individuals is a key problem in mathematical epidemiology. To address this problem, we investigate a two-strain model within a Susceptible-Infected-Recovered (SIR) framework. While classical deterministic theory predicts that the basic reproduction number fully determines selection, we show that stochastic effects play a key role in the dynamics. We discover that stochastic fluctuations can reverse the deterministic advantage even far from the quasi-neutral regime. Further, we find that stochasticity drastically reduces fixation times from years, in the deterministic case, to days. The fixation time is non-linearly proportional to the noise intensity and the distance from the quasi-neutral regime, following a universal rule obtained from a scaling law. The nature of the problem and the equations allow us to interpret the competition as a dynamical evolution around an effective potential, with the potential barrier corresponding to the unstable manifold associated with the coexistence. Even in a stable situation of dominance of one strain, the noise can induce crossings through the potential. We find that the reversal can occur even far from the quasi-neutral regime with significant probability.
\end{abstract}


\end{frontmatter}

\section{Introduction}
Infectious disease dynamics often involve the emergence of multiple pathogen strains through mutations, which subsequently compete for the same pool of susceptible individuals \cite{Suzuki2007}. Multi-strain dynamics are observed across several epidemiological contexts, including COVID-19 \cite{Volz2021}, Influenza \cite{Minayev2009}, Tuberculosis \cite{chin2024}, and Dengue \cite{Aguiar2022}. Therefore, the strain competition is a remarkable problem in mathematical epidemiology.

Different mathematical models have been proposed to study the strain competition in the context of infectious diseases \cite{Andreasen2018,Le2023}. Traditional approaches, based on deterministic evolution rules, hold that a strain with a higher basic reproduction number ($\mathcal{R}_0$) has an ecological advantage over other strains and wins the competition for susceptible individuals \cite{Keeling2008,Driessche2002}. 
However, empirical evidence suggests that competition solely based on the reproduction number cannot fully explain real strain dynamics. For instance, observations in influenza \cite{Bedford2015} and dengue systems \cite{Cummings2004} suggest that strain replacement patterns depend on additional factors such as immunity, spatial structure, and stochastic effects. In terms of mathematical modelling, the dynamical implications of stochastic perturbations remain less explored. In this context, here, we aim to provide a dynamical perspective on how stochastic fluctuations can induce reversals of deterministic selection. In contrast to previous studies restricted to quasi-neutral competition (i.e., similar $\mathcal{R}_0$ for each strain), we demonstrate that stochastic fluctuations can reverse deterministic selection even in strongly asymmetric regimes, and we discover a scaling law that relates fixation times, noise intensity, and distance from the quasi-neutral regime.

Stochastic models offer a more realistic description of epidemic dynamics by accounting for intrinsic and extrinsic sources of randomness that are not present in deterministic formulations \cite{Tome2020,Tome2020b}. These fluctuations may arise from demographic variability associated with discrete infection events, as well as from environmental and behavioural factors \cite{Keeling2008}, such as changes in human activity \cite{Tome2023} or pathogen mutations \cite{Minayev2009}. 
There are clear benefits to using stochastic rather than deterministic models. In the long term, stochastic models can effectively capture the possibility of disease extinction \cite{allen1998,Imran01122013}, while deterministic models typically yield outcomes such as endemic equilibria or a very small number of infected individuals, but not exactly zero.

Different strategies have been proposed to incorporate stochasticity into epidemic models, including demographic noise derived from Markov processes \cite{Stollenwerk2025} and parametric fluctuations in transmission rates \cite{Grenfell2002,Mugnaine2022,Sousa2025}. In the large population limit, such descriptions can often be approximated by stochastic differential equations, providing a continuous framework to analyse fluctuations around deterministic trajectories \cite{AllenBook,GardinerBook,Mario2015}. 

In stochastic modelling, it is crucial to describe the time scales linked to qualitative shifts in system dynamics \cite{Kogan2014}. A widely used concept is the fixation time, which is the time required for the system to reach an absorbing state \cite{Ying2017}. An absorbing state is a state from which the transition rates to all other states are zero \cite{Mario2015}. Consequently, once the system enters this state, it stays there forever. Here, the fixation time represents either the period until the strain becomes dominant in the population or until the disease dies out. Another key concept is the reversal time, defined as the first-passage time at which a stochastic dominance inversion between the competing strains occurs.

Recent advances in strain competition dynamics have highlighted that the strain competition extends beyond traditional deterministic frameworks. The inclusion of cross-immunity and asymmetric temporary immunity periods reveals a complex bifurcation structure, where even when both strains share a practically identical basic reproduction number, heterogeneity in temporary immunity and cross-immunity can alter competitive advantage \cite{Johnston2023}. Fluctuations play a central role in the competition, as noise can shape its long-term outcomes. In finite populations, extinction and persistence are intrinsically probabilistic phenomena, as illustrated by the stochastic SIR model, where epidemic burnout and fade-out events cannot be explained by the deterministic approach \cite{Parsons2024}. Moreover, in quasi-neutral competition, the existence of a deterministic coexistence line shapes the competition under noise perturbations. In this dynamical regime, stochastic fluctuations drive the system to a random walk along this line, leading to the extinction of one strain and the fixation of the other \cite{Kogan2014}. Additionally, the fixation probability may depend on the initial conditions. These mechanisms are related to observation in haploid populations competition \cite{Parsons2007}. Further, when L\'evy noise is considered instead of the white one, the jumps can qualitatively induce transitions in the competitive regimes \cite{Sadki2024}. Together, these studies exhibit how randomness can substantially alter the deterministic competition.

Stochastic formulations of epidemic dynamics can be investigated through several approaches, including Monte Carlo simulations \cite{Tome2022}, cellular automata models \cite{Gabrick2022}, and the direct integration of stochastic differential equations \cite{GardinerBook}. In this work, we adopt the latter framework to investigate strain competition in an SI$_1$I$_2$R model, where $S$ represents the susceptible individuals, $I_{1}$ ($I_2$) the infected by strain 1 (2), and $R$ the recovered. We begin by introducing the deterministic formulation and deriving the conditions for competitive exclusion. As expected, we recover the classical result that the strain with the larger basic reproduction number dominates the dynamics. However, numerical integration reveals that the convergence toward absorbing states occurs on long time scales, typically on the order of years, thereby justifying the inclusion of demographic effects. We then extend the analysis to the stochastic regime, focusing on how noise alters competitive outcomes. By fixing parameters in a region of deterministic dominance, we demonstrate that stochastic fluctuations can induce transitions between dominance states. These transitions significantly affect the temporal organization of strain prevalence, leading to intermittent dominance even for relatively weak noise amplitudes. Furthermore, by computing first-passage times, we show that switching events occur on time scales comparable to fixation times, often on the order of days. This indicates that noise not only induces fixation but also enhances the probability of reversal of dominance, even far from the quasi-neutral regime. To provide a physical interpretation, we introduce an effective potential landscape: far from the critical point, the system is characterized by a pronounced barrier separating the absorbing states, whereas near the critical point, the landscape becomes progressively flatter, facilitating stochastic switching. This behaviour is consistent with noise-driven dynamics observed in quasi-neutral epidemic competition. 

Furthermore, we demonstrate a universal scaling relation that collapses the fixation-time data across different noise intensities, suggesting that the transition between noise-dominated and se\-lection-dominated regimes follows a fundamental scaling law.

This paper is organized as follows. In Section \ref{sec_deterministic}, we introduce the deterministic model and derive the conditions for strain competition. In Section \ref{sec_stochastic}, we present the stochastic formulation. The deterministic dynamics of the log-ratio of populations are analysed in Section \ref{sec_deterministic_q}. In Section \ref{sec_temporal}, we investigate the effects of stochasticity on temporal dominance and switching behaviour. Section \ref{reversal} is devoted to the analysis of first-passage times and reversal probabilities. These results are then interpreted in terms of an effective potential landscape in Section \ref{landscape}. Finally, conclusions and perspectives are presented in Section \ref{conclusions}.

%%%%%%%%%%%%%%%%%%%%%%%%%
%%%%%%%%%%%%%%%%%%%%%%%%%
\section{Deterministic formulation and log-ratio of populations}\label{sec_deterministic}
Here, we study two co-circulating strains in an SI$_1$I$_2$R framework \cite{Johnston2023}, given by 
\begin{eqnarray}
    \frac{dS}{dt} &=& b N - \mu S-\frac{S}{N} (\beta_1 I_1 + \beta_2 I_2), \label{eq_ss}\\
    \frac{dI_1}{dt} &=& \frac{\beta_1}{N} S I_1 - I_1(\gamma_1 + \mu), \label{eq_ii1}\\
    \frac{d I_2}{dt} &=& \frac{\beta_2}{N} S I_2 - I_2(\gamma_2 + \mu),\label{eq_ii2} \\
    \frac{d R}{dt} &=& \gamma_1 I_1 + \gamma_2 I_2 - \mu R, \label{eq_rr1}  
\end{eqnarray}
where $S$ holds susceptible individuals, $I_1$ ($I_2$) represents the infected individual by the first (second) strain with rate $\beta_1$ ($\beta_2$). The recovered rate is $\gamma_1$ ($\gamma_2$), and $b$ ($\mu$) is the birth (death) rate. Finally, $R$ accounts for the recovered individuals. For seek of simplicity, we consider $b=\mu$, which yields $S + I_1 + I_2 + R = N$, where $N$ is the total population \cite{Brugnago2023}. From this constraint, the model can be normalized through the transformation ${\bf x} = {\bf X}/N$, with ${\bf X} = (S,I_1,I_2)$ and ${\bf x}= (s,i_1,i_2)$. Therefore, the model can be rewritten in terms of fractions of individuals, giving rise to
\begin{eqnarray}
    \frac{ds}{dt} &=& \mu - \mu s -s (\beta_1 i_1 + \beta_2 i_2), \label{eq_s}\\
    \frac{d i_1}{dt} &=& \beta_1 s i_1 - i_1 (\gamma_1 - \mu), \label{eq_i1}\\
    \frac{d i_2}{dt} &=& \beta_2 s i_2 - i_2 (\gamma_2 - \mu),\label{eq_i2} 
\end{eqnarray}
where $r = 1 - s - i_1 - i_2$. For biological reasons, we consider $b$, $\mu$, $\beta_{1,2}$, $\gamma_{1,2}$, $s(0)$, $i_{1,2}(0) \geq 0, \ \forall \ t\geq0$.

Equations \eqref{eq_s}-\eqref{eq_i2} admit different equilibrium solutions: $i$) trivial, i.e., $(s,i_1,i_2) = (0,0,0)$. 
$ii$) disease-free equilibrium (DFE), given by $\mathcal{E}_{\rm DFE} \equiv (s^*,i^*_1,i^*_2) = (1,0,0)$; 
$iii$) strain-one (strain-two) dominance, given by $i_1 \neq 0$ and $i_2 = 0$ ($i_2 \neq 0$ and $i_1 = 0$). 
$iv$) simultaneously coexistence of strains one and two: $i_1 \neq 0 $ and $i_2 \neq 0$. 

First, we discuss in detail the stability and properties of the DFE solution. The Jacobian matrix at $\varepsilon_1$ is
\begin{equation}
    J_{\mathcal{E}_{\rm DFE}} =\left(
\begin{array}{ccc}
 -\mu  & -\beta _1 & -\beta _2 \\
 0 & \beta _1-\gamma _1-\mu  & 0 \\
 0 & 0 & \beta _2-\gamma _2-\mu  \\
\end{array}
\right)
\end{equation}
with eigenvalues equal to $\lambda^*_1 = \beta_1 - \gamma_1 - \mu$, related to strain 1, $\lambda^*_2 = \beta_2 - \gamma_2 - \mu$, associated with strain 2, and $\lambda^*_3 = -\mu$. The DFE equilibrium is asymptotically  stable if $\lambda^*_{1} < 0$ and $\lambda^*_{2} < 0$, imposing these conditions, we get $\beta_j/( \gamma_j + \mu) < 1$, where $j = 1,2$. These ratios give the basic reproduction number associated with each strain, defined by 
\begin{align}
\mathcal{R}_1 = \frac{\beta_1}{\gamma_1 + \mu}, 
\qquad
\mathcal{R}_2 = \frac{\beta_2}{\gamma_2 + \mu}.
\end{align}
When $\mathcal{R}_j<1$, the DFE is asymptotically stable; otherwise, it is unstable. Previously, we have denoted the basic reproduction number by $\mathcal{R}_0$; henceforth, we use $\mathcal{R}_j$.

Now, we investigate the boundary equilibrium conditions, i.e., when only one strain is present ($i_1 \neq 0$, while $i_2 = 0$; or  $i_2 \neq 0$, while $i_1 = 0$). From Eqs. \eqref{eq_i1}-\eqref{eq_i2}, we obtain
\begin{equation}
    \widetilde{s}_j = \frac{1}{\mathcal{R}_j}, 
\end{equation}
and from Eq.~\eqref{eq_s}:
\begin{equation}
    \widetilde{i}_j = \frac{\mu}{\beta_j}\left(\mathcal{R}_j - 1\right).
\end{equation}
Therefore, the dominance of strain $j$ is ensured when the equilibrium $\mathcal{E}_j = \left( \widetilde{s}_j, \widetilde{i}_j, 0\right)$ is stable, which exists only when $\mathcal{R}_j > 1$. 

Considering the equilibrium $\mathcal{E}_1$, we obtain three eigenvalues associated with the Jacobian matrix, where one of them is $\widetilde{\lambda}= (\gamma_2 + \mu)(\mathcal{R}_1/\mathcal{R}_2 - 1)$. Therefore, strain-1 dominates over strain-2 only when $\mathcal{R}_1>\mathcal{R}_2$. These conditions ensure the stability of $\mathcal{E}_1$, while $\mathcal{E}_2$ becomes unstable. However, as measure we increase $\mathcal{R}_1$ until $\mathcal{R}_2$  the eigenvalue $\widetilde{\lambda}$ goes to zero and, then, $\mathcal{E}_1$ becomes unstable and $\mathcal{E}_2$ is stable for $\mathcal{R}_2 > \mathcal{R}_1$.  The mechanism behind this exchange of stability is the transcritical bifurcation, and the bifurcation point is $\mathcal{R}_1 = \mathcal{R}_2$.

Another equilibrium point is associated with the coexistence of strains, i.e., $i_1 \neq 0$ and $i_2 \neq 0$. Under these conditions, we obtain two solutions for $s$, that are $s = 1/\mathcal{R}_1$ and $s = 1/\mathcal{R}_2$, holds only for $\mathcal{R}_1 = \mathcal{R}_2$. Therefore, in generic terms, when two strains have different reproduction numbers, coexistence is not possible.

To study in more detail the exchange of dominance, we introduce the log-ratio of populations $q(i_1,i_2;t)$, given by
\begin{equation}
    q(i_1,i_2;t) = {\rm ln} \left(\frac{i_1}{i_2}\right). \label{eq_st_p}
\end{equation}
Instead of individual fractions, $i_1$ and $i_2$, the equivalent form is $q(I_1,I_2;t)$. This log-ratio of populations facilitates the study of competition once we reduce the analysis to one dimension.

The time evolution of Eq.~\eqref{eq_st_p} is given by
\begin{equation}
    \frac{d q}{dt} = s \Delta \beta - \Delta \gamma, \label{deterministic_q}
\end{equation}
where $\Delta\beta = \beta_1 - \beta_2$ and $\Delta\gamma = \gamma_1 - \gamma_2$. 
For a fixed value of $\Delta\gamma$, the sign of $\dot{q}$ determines the 
direction of the competitive dynamics: if $\Delta\beta > 0$, then $\dot{q} > 0$, 
indicating that strain-1 increases its relative prevalence; otherwise 
($\dot{q} < 0$), strain~2 becomes dominant.

The limits of $q$ provide a direct interpretation of the system state. 
When $i_1 \gg i_2$, we obtain $q \rightarrow +\infty$, corresponding to 
dominance of strain~1. Conversely, when $i_2 \gg i_1$, we have 
$q \rightarrow -\infty$, indicating dominance of strain-2. 

%%%%%%%%%%%%%%%%%%%%%%%%%
%%%%%%%%%%%%%%%%%%%%%%%%%
\section{Stochastic formulation}
\label{sec_stochastic}
The dynamics for the deterministic system are already established with the previous results. Now, we extend these results by including stochastic perturbations in the incidence terms, in such a way that the model becomes
\begin{align}
    \frac{dS}{dt} &= b N - \mu S - \left(\beta_1\frac{S I_1}{N} + f(S,I_1)\xi_1\right) \nonumber \\ &- \left(\beta_1\frac{S I_2}{N} + f(S,I_2)\xi_2\right), \label{eq_st_ss}\\
    \frac{dI_1}{dt} &= \left(\beta_1\frac{S I_1}{N} + f(S,I_1)\xi_1\right) - (\gamma_1 + \mu) I_1, \label{eq_st_ii1}\\
    \frac{dI_2}{dt} &= \left(\beta_2\frac{S I_2}{N} + f(S,I_2)\xi_2\right) - (\gamma_2 + \mu) I_2, \label{eq_st_ii2}
\end{align}
where $R = N - S - I_1 - I_2$, the noise acts as a perturbation with amplitude $\xi_j$ on the incidences, modulated by $f(S,I_j) = \sqrt{\beta_jSI_j/N}$. 
Under these assumptions, the It\^o SDEs for $I_j$ are
\begin{equation}
    d I_j = \left[\beta_j S \frac{I_j}{N} - (\gamma_j + \mu) I_j\right]dt + \xi_j \sqrt{\beta_j S \frac{I_j}{N}} dW_j, \label{ito_i}
\end{equation}
where $W_j$ is a Wiener independent noise. 

The stochastic correspondence of $dq/dt$, now becomes
\begin{align}
    dq &= \left[s\Delta\beta - \Delta \gamma - \frac{1}{2} \left(\xi_1^2 \frac{\beta_1 s}{i_1} - \xi_2^2 \frac{\beta_2 s}{i_2} \right)\right]  dt \nonumber \\
    &+ \xi_1 \sqrt{\frac{\beta_1 s}{i_1}} dW_1 - \xi_2 \sqrt{\frac{\beta_2 s}{i_2}}d W_2, \label{stochastic_q}
\end{align}
when $\xi_1 = \xi_2 = 0$ we recover  Eq. \eqref{deterministic_q}. It is worth noting that the noise term in Eq. \eqref{stochastic_q} scales as $1/\sqrt{i_j}$, which implies that fluctuations become singular as the system approaches the extinction of either strain ($i_j\rightarrow0$).  Further, the stochastic dynamics of the system can be interpreted as a random evolution of $q(i_1,i_2;t)$ between the absorbing states $q(i_1,i_2;t) \rightarrow \pm \infty$, which correspond to fixation of one of the strains.

Equation~\eqref{stochastic_q} can be written in the form of a Langevin equation \cite{GardinerBook,Mario2015}:
\begin{equation}
dq = A(q)\,dt + B(q)\,dW,
\label{langevin}
\end{equation}
where $A(q)$ represents the deterministic drift and $B(q)$ the noise
intensity. Both coefficients are implicitly dependent on $q$, once $i_j$ can be rewritten in terms of $q$.
If the drift term can be expressed as the gradient of a potential function, i.e., $A(q) = - \partial_q U(q)$, the stochastic dynamics can be interpreted as diffusion in an effective potential landscape, with
\begin{equation}
dq = -\frac{\partial U(q)}{\partial q}\,dt + B(q)\,dW.
\label{potential}
\end{equation}

The numerical integration of the stochastic system is performed through the Euler--Maruyama method with a fixed step size of $h=1/(2*365)$. In our simulations, we use $N=10^5$ individuals.
%%%%%%%%%%%%%%%%%%%%%%%%%
%%%%%%%%%%%%%%%%%%%%%%%%%
\section{Deterministic solution of log-ratio of populations}\label{sec_deterministic_q}
\begin{figure}
    \centering
    \includegraphics[scale=0.7]{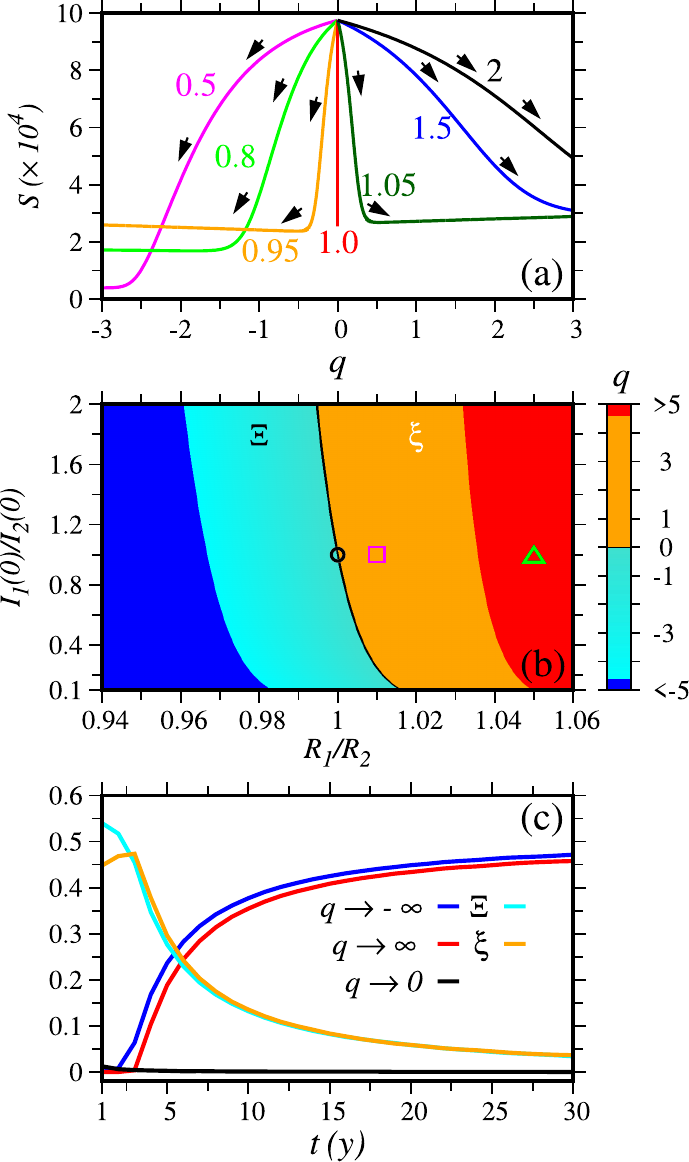}
    \caption{(a) Evolution of Susceptible individuals ($S$) as a function of $q$ for distinct values of the ratio between the basic reproduction numbers --- $\mathcal{R}_1 / \mathcal{R}_2$. In this panel, the magenta, green, orange, red, dark-green, blue, and black curves correspond to $\mathcal{R}_1 / \mathcal{R}_2 = 0.5$, 0.8, 0.95, 1, 1.05, 1.5, and 2, respectively. The initial conditions are $I_1(0)=I_2(0)=0.025N$. (b) Parameter plane showing the dependence of $q$ in the pairs $I_1(0)/I_2(0) \, \times \, \mathcal{R}_1 / \mathcal{R}_2$. We fix $I_2(0)=0.025N$. The results of panels (a) and (b) are for five years of integration. (c) display the size of the corresponding basins exhibited in panel (b) as a function of time in years. }
    \label{fig1}
\end{figure}

Figure \ref{fig1}(a) shows the evolution of Susceptible individuals ($S$) as a function of $q$, for a life expectancy rate of $1/\mu = (70*365)$, and a recovery rate of $1/\gamma = 7$ --- the time unit is days. In this result, we consider five years of integration and $I_1(0) = I_2(0) = 0.025N$. In the region of two dominance strains, the smaller $\mathcal{R}_1 / \mathcal{R}_2$, the smaller is the long-term population $S$, and the smaller is $q$, which tends to $-\infty$. In this case, the reduction of $S$ occurs mainly due to infections caused by the second strain. A critical region is for $\mathcal{R}_1 / \mathcal{R}_2 = 1$, where there is an equilibrium between the dominance and the value of $q$ is zero. Positive values of $q$, appear for the dominance of strain one ($\mathcal{R}_1 / \mathcal{R}_2 > 1$). 

The results shown in Fig. \ref{fig1}(a) are restricted to one initial condition and fixed values of $\mathcal{R}_1 / \mathcal{R}_2$. In Fig. \ref{fig1}(b), we explore a broader situation by varying $I_1(0)/I_2(0)$ and $\mathcal{R}_1 / \mathcal{R}_2$, where the colour scale shows $q$. This outcome is also computed for five years of integration, and $I_2(0) = 0.025N$. We note five distinct regions in the basin of attraction, where blue and red identify the absorbing states, cyan and orange delimit intermediate regions, and black shows the coexistence. The results in the transient dynamics reveal that the competition depends on a non-linear relationship between initial conditions and $\mathcal{R}_1/\mathcal{R}_2$.

Each basin in Fig. \ref{fig1}(b) has a given size in relation to the parameter plane, and the evolution of this measure is shown in Fig. \ref{fig1}(c). For five years of integration, we are nearly the inflection point where the regions $\Xi$ and $\xi$ decay in size and the absorbing basins (blue and red) increase in size. The steady state is reached within 30 years of integration, indicating a long time to fixation. Furthermore, the size of the coexistence basin (black region) is practically null throughout the integration time, particularly after the first two years.  This outcome suggests that coexistence for longer periods (greater than 2 years) is practically impossible to achieve due to the particularities of these states: specific ratios of initial conditions and exact values of the basic reproduction numbers. 

%%%%%%%%%%%%%%%%%%%%%%%%%
%%%%%%%%%%%%%%%%%%%%%%%%%
\section{Noise-induced reversal time occupancy and crossings over the barrier}\label{sec_temporal}
Considering the three points highlighted in Fig. \ref{fig1}(b) by the black circle ($\mathcal{R}_1 / \mathcal{R}_2 = 1$), purple square ($\mathcal{R}_1 / \mathcal{R}_2 = 1.01$), and green triangle  ($\mathcal{R}_1 / \mathcal{R}_2 = 1.05$), all of them with $I_1(0)/I_2(0) = 1$ we now explore the effects of $\xi_j$ in the dynamics.  

To explore the effects of the noise, we integrate the system over an effective time window of length $T_{\rm eff}$ --- which means a region before the extinction of one strain. For each realization, we compute the temporal occupation fractions $\chi_1$, $\chi_2$, and $\chi_{\rm c}$ for each state, i.e., the fraction of time spent in the dominance region of strain 1, strain 2, and in the coexistence region, respectively. These fractions are, respectively, computed when $q> q_{\rm c} = 0.02$, $q<-q_{\rm c}$, and $- q_{\rm c} \leq q \leq q_{\rm c}$.  Formally, the  averages of these fractions are defined as
\begin{align}
    \langle \chi_1 \rangle &= \frac{1}{N_{\rm r}} \sum_{r=1}^{r=N_r} \frac{1}{T_{\rm eff}} \int_0^{T_{\rm eff}} \, H(q_r(t) - q_{\rm c}) \, dt, \\
     \langle \chi_2 \rangle &= \frac{1}{N_{\rm r}} \sum_{r=1}^{r=N_r} \frac{1}{T_{\rm eff}} \int_0^{T_{\rm eff}} \, H(- q_r(t) - q_{\rm c}) \, dt, \\
     \langle \chi_{\rm c} \rangle &= \frac{1}{N_{\rm r}} \sum_{r=1}^{r=N_r} \frac{1}{T_{\rm eff}} \int_0^{T_{\rm eff}} \, H(q_c - |q_r(t)|) \, dt,
\end{align}
where $H(x)$ is the Heaviside function defined according to: $H(x) = 0$ if $x<0$, and $H(x) = 1$ if $x\geq 1$; $N_r$ is the number of realisations, equal to 500 independent simulations, and $\langle\chi_1\rangle + \langle\chi_2\rangle + \langle\chi_{\rm c}\rangle = 1$.

\begin{figure*}
    \centering
    \includegraphics[scale=0.6]{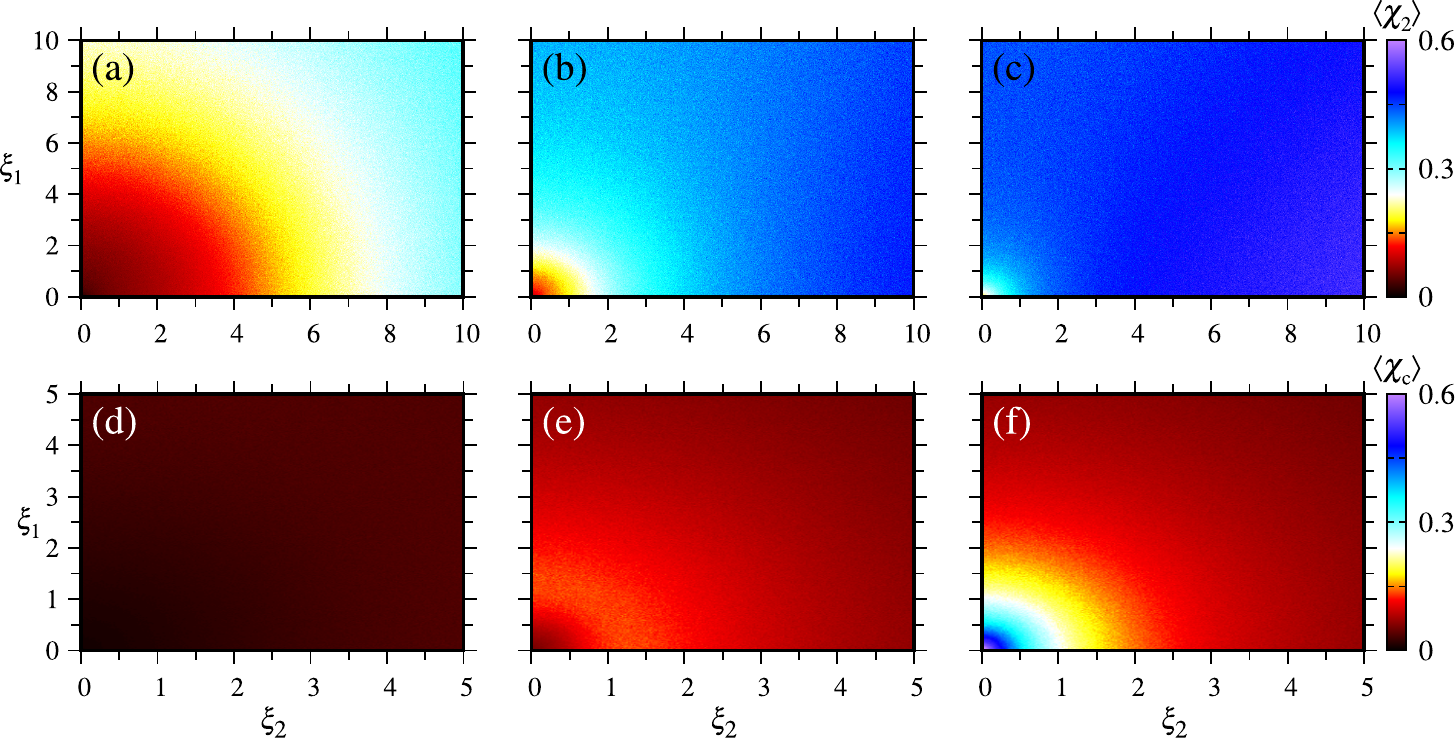}
    \caption{Temporal occupation fractions $\langle\chi_2\rangle$ (panels (a), (b), and (c)),  and $\langle\chi_{\rm c}\rangle$ (panels (d), (e), and (f)), for the dominance of strain 2 and coexistent states as a function of the pairs $(\xi_2,\xi_1)$. In (a) and (d): $\mathcal{R}_1/\mathcal{R}_2 = 1.05$; in (b) and (e): $\mathcal{R}_1/\mathcal{R}_2 = 1.01$; in (c) and (f): $\mathcal{R}_1/\mathcal{R}_2 = 1.0$.}
    \label{fig2}
\end{figure*}

Figures \ref{fig2}(a), \ref{fig2}(b), and \ref{fig2}(c) displays the parameter planes $(\xi_2,\xi_1)$ as a function of  $\langle\chi_2\rangle$ in the colour scale, for  $\mathcal{R}_1 / \mathcal{R}_2 = 1.05$, $\mathcal{R}_1 / \mathcal{R}_2 = 1.01$, and $\mathcal{R}_1 / \mathcal{R}_2 = 1.0$, respectively. Note that only in the last case is the system in coexistence; otherwise, it is in deterministic dominance of strain 1. 

For $\mathcal{R}_1/\mathcal{R}_2 = 1.05$, even for low levels of noise, an exchange of dominance appears for a short time of occupancy, typically less than 25\% of the simulation time. However, a large amplitude of noise combinations (i.e., greater than 6) drives the system to spend more than 40\% of the time in state 2 dominance. In this way, these results highlight that dominance shifts over a significant time window, but do not indicate which strain will ultimately dominate. 

As we approach the critical point, the scenario changes (Fig. \ref{fig2}(b)). Now, lower noise amplitudes are required to change the times of occupancy. In general, red tones ($<$ 15\%) appear in Fig. \ref{fig2}(b) only for noise amplitudes lower than 2, and a significant part of the parameter plane shows $\langle\chi_2\rangle>0.3$. Remarkably, for certain pairs of noise amplitude, the system can reach more than 50\% of the effective epidemic time in the strain 2 dominance, highlighting the noise-driven reversal competition. Therefore, in this case, certain levels of noise can induce bifurcation, bringing the system from strain 1 dominance to strain 2 for longer epidemic times. 

In the critical point (Fig. \ref{fig2}(c)), several regions of the parameter plane display the transitions from strain 1 dominance to 2, and vice-versa. In other words, we now have a mixed dominance, where each strain occupies around 50\% of the epidemic period. This result shows that at the critical point, the system is very sensitive to stochastic fluctuations, and only small noise levels favour maintaining strain-1 dominance for longer times.

As observed, the noise drives the system between the two states, even reversal of dominance for considerable periods; however, the system also spends time in the coexistence region, as displayed in Figs. \ref{fig2}(d), \ref{fig2}(e), and \ref{fig2}(f). Far from the critical point, the system practically does not spend time in the coexistence band, but just crosses this barrier. As we approach the equilibrium, the system starts to spend longer times in the barrier, especially for the level of noise. These outcomes highlight that coexistence is very sensitive and appears to be a barrier separating the two dominances, thereby being highly unstable to small perturbations. 

The mean occupancy time quantifies how the system spends time in one state and in another. However, this measure does not show how many times, on average, the system crosses the barrier. To investigate the noise-induced switching between dominance states, we define $\eta_r$ as the total number of crossings between the states separated by $|q_r| \le q_c$, for 500 realizations.

\begin{figure}
    \centering
    \includegraphics[scale=0.7]{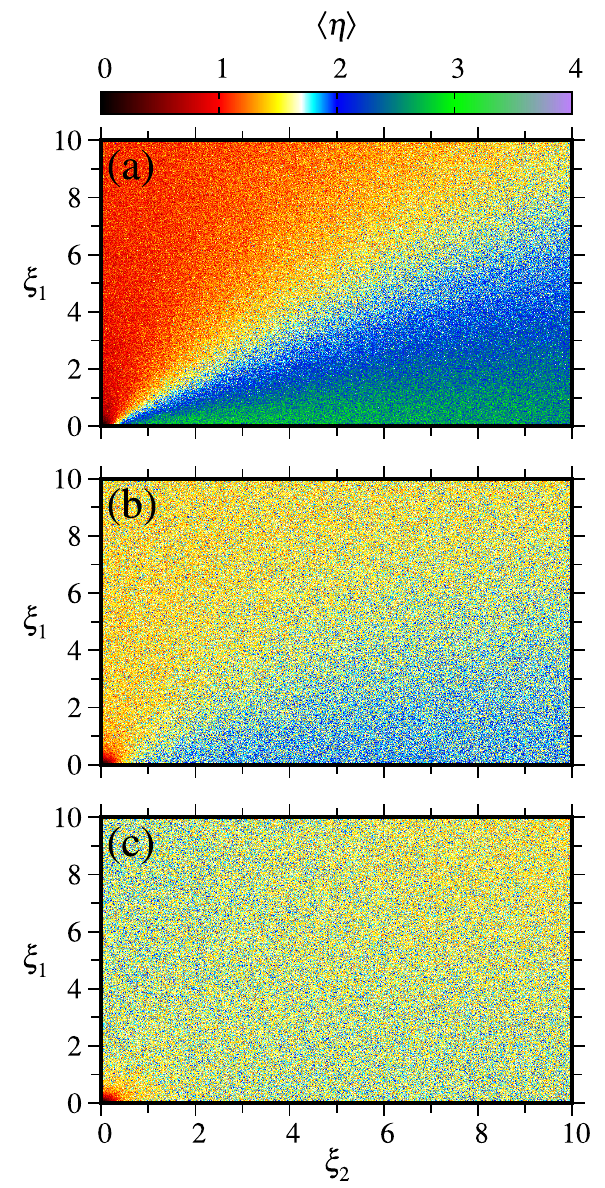}
    \caption{Average of switching times $\langle\eta\rangle$ for (a) $\mathcal{R}_1/\mathcal{R}_2 = 1.05$, (b) $\mathcal{R}_1/\mathcal{R}_2 = 1.01$, and (c) $\mathcal{R}_1/\mathcal{R}_2 = 1.0$.}
    \label{fig3}
\end{figure}

Far from the critical point, the behaviour of $\langle \eta \rangle$ exhibits distinct dynamical regimes for each pair $(\xi_2,\xi_1)$, as shown in Fig. \ref{fig3}(c). In the region where $\xi_1 > \xi_2$, which favours strain 1 dominance, only a few switching events are observed, mostly indicated by red tones in Fig. \ref{fig3}(a). In this regime, the deterministic forces suppress the noise-induced transitions. 

As $\xi_1 \sim \xi_2$, a transition region emerges (white and blue tones), where approximately two dominance exchanges occur on average. This reflects a reduction in the effective asymmetry between the competing basins, allowing stochastic fluctuations to induce more frequent crossings. 

In the regime $\xi_2 > \xi_1$, switching becomes more frequent, particularly for $\xi_1 < 2$. Here, the weaker deterministic forces combined with stronger relative noise enhance the rate of basin-to-basin transitions. Overall, these results reveal a structured landscape of noise-induced switching, in which the interplay between deterministic asymmetry and stochastic forcing determines the frequency of dominance exchanges.

As the system approaches the near-critical regime (Fig. \ref{fig3}(b)), the strong red tones observed in panel (a) gradually disappear, giving rise to predominantly yellow and blue regions. This homogenization of colours indicates a reduction in the spatial variance of $\langle \eta \rangle$ across the $(\xi_2,\xi_1)$ plane. In other words, the influence of noise becomes more symmetric, producing similar switching statistics over a broader parameter range. Considering the pairs where $\langle\chi_2\rangle \sim 0.5$ from Fig. \ref{fig2}(b),  we observe that the system crosses the barrier around  1 and 2 times. Suggesting that when few crossings occur, the dominance reversal occurs. 

At the critical point (Fig. \ref{fig3}(c)), most parameter pairs $(\xi_2,\xi_1)$ yield $\langle \eta \rangle \approx 1.6$ (mean across $(\xi_2,\xi_1)$ plane). In this regime, neither strain possesses a deterministic advantage, and the system operates near a quasi-neutral manifold. The competition between the two dominance basins becomes effectively balanced, which is consistent with the observation that the average value of $\langle \chi_2 \rangle$ over $(\xi_2,\xi_1)$ is approximately $47\%$, close to the symmetric value of $50\%$. 

%%%%%%%%%%%%%%%%%%%%%%%%%
%%%%%%%%%%%%%%%%%%%%%%%%%
\section{Noise-driven  reversal and scaling law}\label{reversal}
\begin{figure}
    \centering
    \includegraphics[scale=0.7]{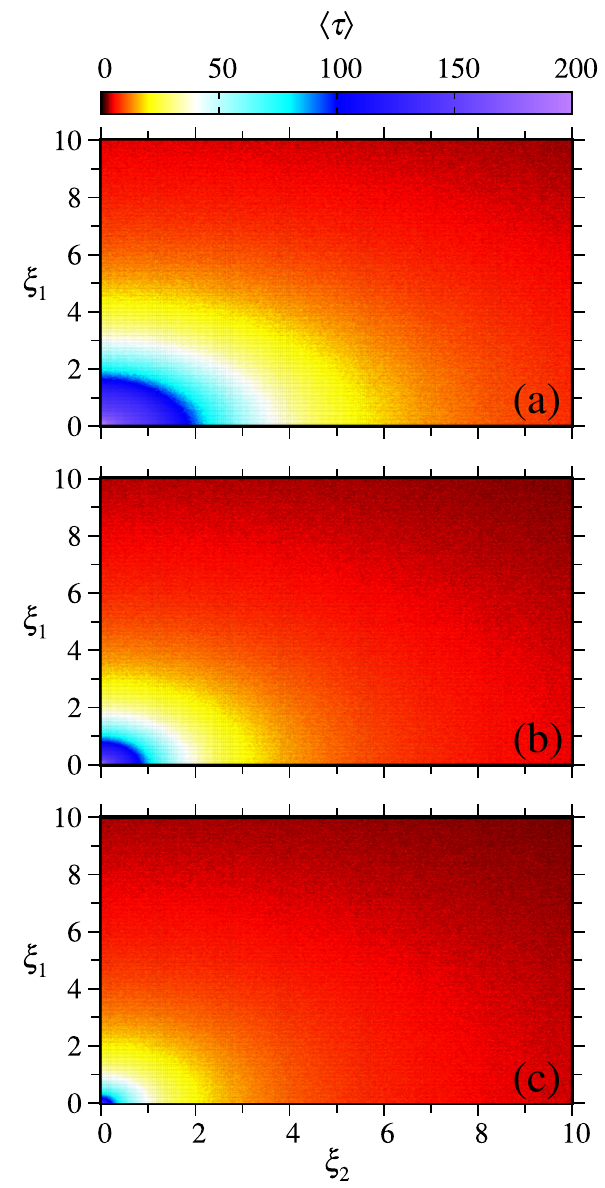}
    \caption{Average of first passage times $\langle\tau\rangle$ for (a) $\mathcal{R}_1/\mathcal{R}_2 = 1.05$, (b) $\mathcal{R}_1/\mathcal{R}_2 = 1.01$, and (c) $\mathcal{R}_1/\mathcal{R}_2 = 1.0$.}
    \label{fig4}
\end{figure}

Now, we compute the average first-time transition $\langle\tau\rangle$ for the pair $(\xi_2,\xi_1)$, considering the same ensemble of simulations. The results for  $\mathcal{R}_1/\mathcal{R}_2 = 1.05$, $\mathcal{R}_1/\mathcal{R}_2 = 1.01$, and $\mathcal{R}_1/\mathcal{R}_2 = 1.0$ are displayed in Fig. \ref{fig4} in the panels (a), (b), and (c), respectively.

Figure \ref{fig4}(a) displays distinct regimes of mean first-passage times depending on the pair $(\xi_2,\xi_1)$. As expected, for low noise intensities (typically less than 1), the system spends a long time ($\sim 200$ days) before the first passage. This effect of long times to occur transitions reflects into $\langle\eta\rangle \sim 1$ (Fig. \ref{fig3}(a)) and $\langle\chi_2\rangle \sim 0$ (Fig. \ref{fig2}(a)). As the noise amplitude increases, the times of first passage decrease. They grow in the parameter plane as circular structures, with boundaries defined by a nonlinear relationship between the amplitudes.  In a region where $\xi_{1,2}>3$, we obtain $\langle\tau\rangle<50$ days. A few days (less than 10) for occurring transitions are observed only for stronger noise. This suggests that for a region deterministically dominated by strain 1, it is necessary to have high amplitudes of noise in both strains to cause short-time transitions of dominance. However, it is not necessary for an exchange of dominance, but changes switching between them, as previously noted in Fig. \ref{fig3}(a) and Fig. \ref{fig2}(a).

As we approach the critical point, lower values of $\langle\tau\rangle$ are obtained for lower values of $\xi_{1,2}$ (Fig. \ref{fig4}(b)). Now, the transition from one attractor-like state to another occurs faster because the system is more sensitive. Finally, when we reach the critical point, the parameter plane $\xi_1 \times \xi_2$ mostly presents lower values of $\langle\tau\rangle$. This indicates that the first-crossing state occurs earlier for most amplitude-noise combinations. High values of $\langle\tau\rangle$, typically around 100 days, are only reached for $\xi_{1}<0.5$ and $\xi_2<0.5$. 

Another central quantity in stochastic competition dynamics is the fixation time $\langle\tau\rangle_\textrm{fix}$. In contrast to the first-passage times discussed in Fig. \ref{fig4}, the fixation time measures the irreversible selection of a single strain.

\begin{figure}[t]
    \centering
    \includegraphics[scale=0.7]{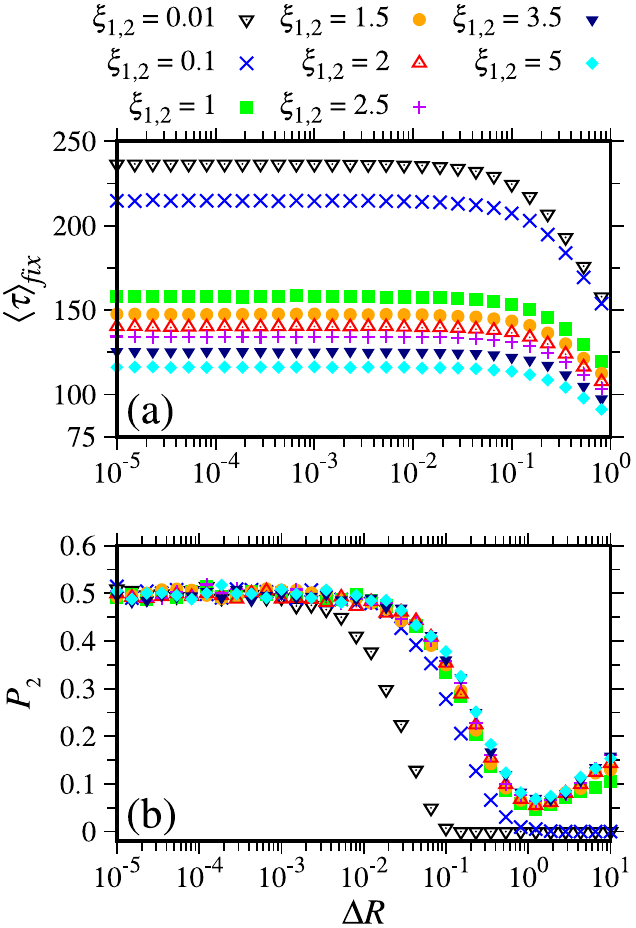}
    \caption{Mean fixation time $\langle\tau\rangle_\textrm{fix}$ as a function of the control parameter $\Delta \mathcal{R} = \mathcal{R}_1 - \mathcal{R}_2$ for different noise intensities $\xi_{1,2}$. We consider 5{,}000 independent simulations and $\mathcal{R}_1 \gtrsim \mathcal{R}_2$.}
    \label{fig5}
\end{figure}

To estimate $\langle\tau\rangle_\textrm{fix}$, we perform an ensemble average over $5,000$ independent realizations for fixed noise amplitudes $\xi_{1,2}$. We, then, introduce the control parameter $\Delta \mathcal{R} = \mathcal{R}_1 - \mathcal{R}_2$, with $\mathcal{R}_1 \gtrsim \mathcal{R}_2$, such that the deterministic dynamics favours strain~1 (Fig. \ref{fig5}(a)). Due to deterministic effects, we know that strain-1 always wins in this setup. However, in the presence of stochastic fluctuations, sometimes this result is not true. To measure this, we compute the probability of strain-2 winning the competition ($P_2$), i.e., the probability of reversal (see Fig. \ref{fig5}(b)).

Figure \ref{fig5}(a) shows $\langle\tau\rangle_\textrm{fix}$ as a function of $\Delta \mathcal{R}$ for different noise intensities. In this outcome, two distinct regimes emerge. Close to the critical point ($\Delta \mathcal{R} \lesssim 0.1$), the fixation time exhibits a plateau, remaining approximately constant and controlled solely by the noise amplitude. This constant is $\langle\tau\rangle_0 \equiv \langle\tau\rangle_\textrm{fix} (\Delta \mathcal{R} \rightarrow 0)$. In this quasi-neutral regime, deterministic selection is weak and stochastic fluctuations dominate the dynamics. As a result, fixation occurs through noise-driven diffusion along the critical point.

For larger asymmetry ($\Delta \mathcal{R} \gtrsim 10^{-1}$), the fixation time decays proportionally to $\Delta \mathcal{R}$,  indicating that deterministic selection progressively dominates the dynamics. This trend reflects the competition between drift and diffusion, and stronger noise substantially reduces fixation time.

An additional important feature revealed in Fig. \ref{fig5}(b) is the presence of noise-induced selection reversal. Although $\mathcal{R}_1 > \mathcal{R}_2$ ensures that $q \to +\infty$ is the only deterministic attractor, stochastic fluctuations allow trajectories to reach the opposite absorbing state $q \to -\infty$ with non-zero probability. This effect persists even far from criticality ($\Delta {\mathcal{R}} > 1$).

Near the critical point ($\Delta \mathcal{R} \lesssim 10^{-3}$), the fixation probability becomes approximately 50\%. This result demonstrates that, in the quasi-neutral regime, the outcome of competition is intrinsically stochastic and cannot be predicted solely from deterministic arguments.

\begin{figure}
    \centering
    \includegraphics[scale=0.7]{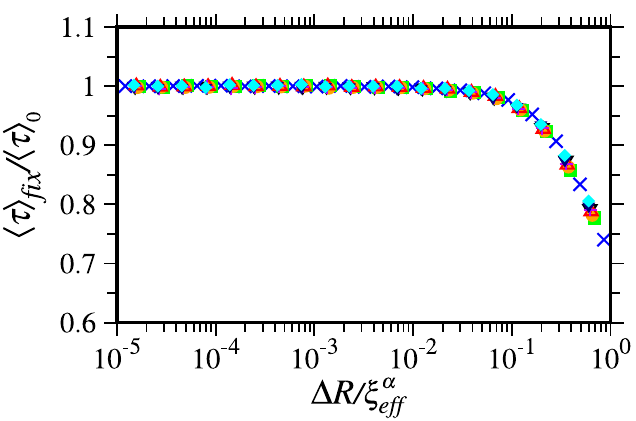}
    \caption{Universal scaling of the mean fixation time. By rescaling the variables as $\langle\tau\rangle_{\mathrm{fix}}/\langle\tau\rangle_0$ and $\Delta \mathcal{R}/\xi_{\mathrm{eff}}^{\alpha}$, all curves corresponding to different noise intensities collapse onto a single master curve, with exponent $\alpha = 0.1$. This result indicates an invariant balance between deterministic drift and stochastic diffusion that governs fixation dynamics across parameter regimes. The breakdown of the collapse for weak noise and strong asymmetry highlights the limits of the scaling law.}
    \label{fig6}
\end{figure}

Figure \ref{fig5}(a) reveals a scaling relationship between the relevant variables, which remains robust across a wide range of noise intensities. By rescaling the axes by $\langle\tau\rangle_{\mathrm{fix}}/\langle\tau\rangle_0$ and $\Delta \mathcal{R}/\xi_{\mathrm{eff}}^{\alpha}$, with $\xi_{\mathrm{eff}}^2 = \xi_1^2 + \xi_2^2$, we observe a collapse of all curves onto a single master curve (Fig. \ref{fig6}). The optimal collapse is achieved for $\alpha = 0.1$.

For very weak noise levels (typically $\xi_{1,2} < 10^{-3}$), deviations from the scaling behaviour emerge, indicating a breakdown of universality. Although the scaling law remains robust for symmetric and moderately asymmetric noise, deviations are observed under extreme asymmetry (e.g., $\xi_1\gg\xi_2$). This breakdown indicates the emergence of a noise-induced bias (spurious drift) that breaks the symmetry of the effective potential, a feature that a global scaling factor cannot fully account for. This limitation defines the validity domain of the proposed universality, which holds as long as the stochastic pressures on both strains are of comparable orders. 

This data collapse is a signature of universality in stochastic systems, suggesting that the transition from a noise-dominated plateau to a selection-driven regime is governed by a fundamental scaling function $\mathcal{F} (\Delta \mathcal{R}/\xi_{\rm eff}^\alpha)$ \cite{Denis2026,Denis2025}. Such an observation implies that, despite the nonlinearity of the epidemic model and the nature of the fluctuations, the underlying competition dynamics follow a universal principle: the ratio between the deterministic drift and the stochastic strength uniquely determines the fixation timescale.
%%%%%%%%%%%%%%%%%%%%%%%%%%
%%%%%%%%%%%%%%%%%%%%%%%%%%
\section{Effective landscape interpretation}\label{landscape}
\begin{figure}
    \centering
    \includegraphics[scale=0.32]{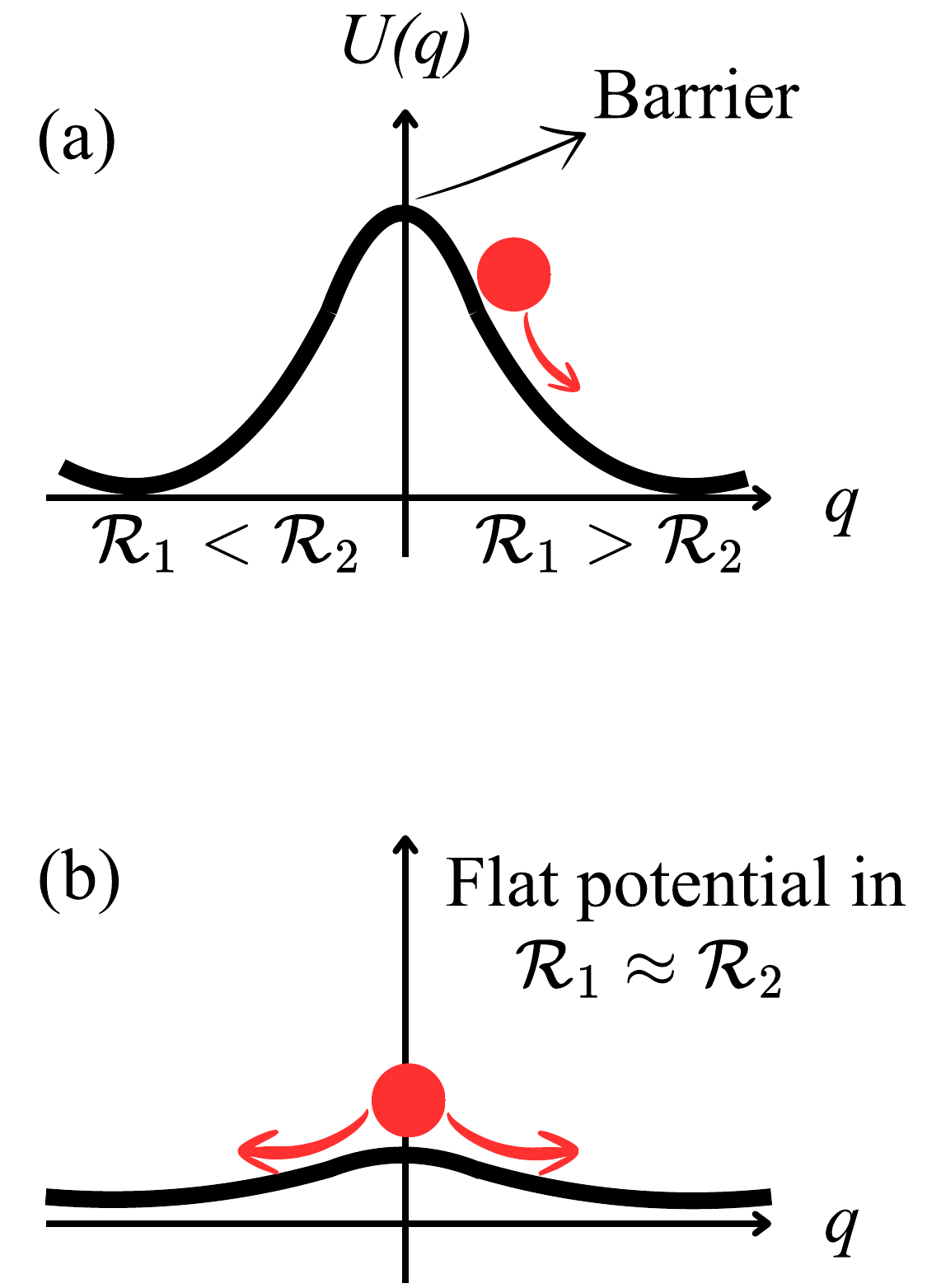}
    \caption{Schematic representation of the effective potential $U(q)$ governing
the stochastic competition dynamics. 
(a) For $\mathcal{R}_1 \neq \mathcal{R}_2$, the potential is tilted,
creating a barrier that separates the dynamics toward the dominant
strain. 
(b) For $\mathcal{R}_1 \approx \mathcal{R}_2$, the potential becomes
nearly flat, leading to fluctuation-dominated dynamics around the
coexistence region. }
    \label{fig7}
\end{figure}

To interpret these results, from a physical perspective, we consider the effective potential formulation introduced in Eq. \eqref{potential} and, through our results, we infer a landscape $U(q)$, where absorbing states correspond to the limits $q \to \pm \infty$, represented in Fig. \ref{fig7}.

Far from the critical point (Fig.~\ref{fig7}(a)), the potential has an effective barrier located at $q=0$. This barrier separates the two dominant states. This conceptualization maps our epidemic competition model onto the classic Kramers escape problem \cite{kramers1940}, where the mean fixation time is analogous to the noise-driven activation time required to cross an energy barrier.  Considering a start in $\mathcal{R}_1>\mathcal{R}_2$, a noise-induced transition can activate a reversal dominance for significant values of noise intensity. This explains the first panels of Figs. \ref{fig2}, \ref{fig3}, \ref{fig4}, where the occupancy time in state 2 is lower for lower noise amplitude, few crossings occur for small values of noise, and first passage times are shorter only for high levels of noise. This is also connected with particular cases in Fig. \ref{fig5} where, although there are crossings, the reversal probability is lower. 

In contrast, near the critical point (Fig. \ref{fig7}(b)), the potential becomes progressively flatter as we have $\Delta \mathcal{R} \to 0$. The deterministic drift vanishes (see Eq. \eqref{stochastic_q}) and the system evolves along a quasi-neutral direction. In this regime, even weak noise is sufficient to induce transitions between dominance states, leading to frequent switching and long residence times near $q \approx 0$, as reported in Fig. \ref{fig3}(c) and almost half of the occupancy time being in one state (Fig. \ref{fig2}(c)). The dynamics are therefore dominated by diffusion, and fixation emerges as a first-exit process from a nearly flat landscape. This diffusive phenomenon brings about uncertainty about the strain that will be absorbed. The observed data collapse suggests that the effective barrier height $\Delta U$ exhibits a nonlinear dependence on noise intensity. In particular, the rescaling $\Delta \mathcal{R} \rightarrow \Delta \mathcal{R}/\xi_{\mathrm{eff}}^{\alpha}$ preserves the balance between deterministic drift and stochastic diffusion, indicating an underlying invariance of the effective dynamics.

Interestingly, the proposed scaling law $\eta = \Delta \mathcal{R}/\xi_{\rm eff}^\alpha$ exhibits a breakdown in regimes of high noise asymmetry (e.g., $\xi_1 = 1$ and $\xi_2 \ll1$). This deviation highlights the emergence of a noise-induced bias. While $\xi_{\rm eff}$ accounts for the total diffusive power, it does not capture the spurious drift introduced by the imbalance in stochastic fluctuations (see Eq. \eqref{stochastic_q}). In such cases, the effective potential is not only flattened but also undergoes a symmetry-breaking deformation that cannot be captured by a simple global scaling factor. This defines the validity domain of our universal scaling, which remains robust as long as the noise intensities are of the same order of magnitude.

This picture provides a unified interpretation of the results reported in Figs.~\ref{fig2}---\ref{fig5}. The plateau in $\langle\tau\rangle_\textrm{fix}$ corresponds to diffusion-limited dynamics in a flat potential, while the power-law regime reflects drift-dominated escape in a curved landscape. Moreover, the reduction of fixation times from years (deterministic case, Fig.~\ref{fig1}(c)) to days (stochastic case, Fig.~\ref{fig5}) highlights the crucial role of noise in accelerating selection.

Overall, our results demonstrate that stochasticity not only perturbs the deterministic outcome but fundamentally reshapes the competitive dynamics, enabling switching between dominance states and controlling the timescale of fixation.
%%%%%%%%%%%%%%%%%%%%%%%%%
%%%%%%%%%%%%%%%%%%%%%%%%%
\section{Conclusions} \label{conclusions}
In this work, we have analysed stochastic strain competition in an SI$_1$I$_2$R epidemic model, focusing on the role of fluctuations in shaping competitive results. We derive deterministic results that predict a strict hierarchy governed by the basic reproduction numbers, $\mathcal{R}_1$ and $\mathcal{R}_2$. However, our results show that this hierarchy is not robust under stochastic perturbations --- and deterministic dominance can be reversed under perturbations. Additionally, we show a scaling law relating the fixation time to the noise amplitude and the distance from the quasi-neutral regime.

In the deterministic regime, the strain with the largest reproduction number always dominates, and coexistence is unstable, with a very small basin of attraction. In contrast, when stochastic effects are included, the dynamics of the competition process are fundamentally altered. In particular, fixation times become shorter (reduction from years to days), and the time occupancy of each strain is substantially modulated by noise intensity. 

The main novelty of this work is to show the possibility of stochastic reversal of deterministic selection. Even when $\mathcal{R}_1 > \mathcal{R}_2$, the inferior strain can reach fixation with significant probability. This effect arises from the interplay between nonlinear infection dynamics and stochastic dynamics, which modifies both the effective drift and fluctuations in the system.

This mechanism differs qualitatively from previously studied quasi-neutral competition scenarios, where stochastic effects dominate only in the limit $\mathcal{R}_1 \sim \mathcal{R}_2$. On the other hand, our results show that noise-induced reversals can occur far from the quasi-neutral regime, where deterministic selection is expected to be robust. In this way, we show that noise can fundamentally reshape the competitive landscape.

Moreover, we show that fixation times and reversal probabilities present a non-trivial dependence on $\mathcal{R}_1 - \mathcal{R}_2$, with $\mathcal{R}_1 > \mathcal{R}_2$, and noise amplitude. We have two regimes: the first, near the critical point, is a plateau; the second, far from the critical point, is a power-law regime. On the other hand, in the limit $\mathcal{R}_1 \rightarrow \mathcal{R}_2$ we measure that the probability of reversal tends to 50\%.

Beyond the noise-induced reversal, we identify a scaling law characterized by a universal exponent, revealing a nonlinear relationship among fixation time, noise intensity, and the distance from the critical regime. We further delineate the validity domain of this scaling behaviour. In particular, deviations emerge in the weak-noise limit, where the dynamics approach the deterministic regime, and in strongly asymmetric noise configurations, where one strain dominates the stochastic forcing.

We explain these findings using an effective potential landscape, in which the barrier corresponds to the coexistence state separating the two absorbing solutions. As we approach the critical point, the potential becomes flat, and the system performs a random walk around the coexistence line. Furthermore, even far from the critical point, noise induces crossings through the barrier. 

These findings provide a minimal dynamical explanation for strain replacement patterns that cannot be explained purely based on the basic reproduction numbers. In particular, our outcomes suggest that stochastic effects can control the direction and timescale of strain competition.  

In future works, we aim to derive analytical predictions of fixation probabilities and times. Additionally, we plan to extend this framework to broader scenarios, including metapopulations and heterogeneous environments.

%%%%%%%%%%%%%%%%%%%%%%%%%
%%%%%%%%%%%%%%%%%%%%%%%%%
\section*{Acknowledgements}

The authors thank the financial support from the Brazilian Federal Agencies (CNPq), under grant No. 302665/2017-0,  CAPES, and the S\~ao Paulo Research
Foundation (FAPESP, Brazil), under grants No. 2024/05700-5. E.C.G. acknowledges financial support from FAPESP under grants Nos.  2025/02318-5, and 2025/24097-0. A.L.M. acknowledges financial support from FAPESP under grant 2025/05453-0. E.K.L. thanks the partial support of the CNPq under grant No. 301715/2022-0.

\bibliographystyle{elsarticle-num}
\bibliography{references}

\end{document}